\documentclass[12pt]{iopart}
\usepackage{iopams}
\begin{document}

\title{Ising tricriticality and the dilute A$_3$ model.}
\author{Katherine A Seaton}
\address{School of Mathematical and Statistical Sciences, La Trobe University,
Victoria 3086, Australia}

\begin{abstract}Some universal amplitude ratios appropriate to the $\phi_{2,1}$
perturbation of the $c=\case{7}{10}$ minimal field theory,
the subleading magnetic perturbation of the tricritical Ising model, are explicitly 
demonstrated in the dilute A$_3$ model, in regime 1. 
\end{abstract}
%
%
\section{Introduction}\label{intro}
While the idea that universality should be described not only by
referring to critical exponents but also to universal amplitude ratios
is not new, it continues to attract interest.  Universal ratios are constructed in such a way as to cause cancellation of any
metric factors associated with the particular realization of a universality class being considered. Thus comparison of
these ratios, say from a field theory and a lattice model, or as measured in an experimental system, increases
our understanding of universality. Alternatively, they may be considered to have predictive value.

Since the time of the comprehensive review of universal critical-point
amplitude ratios in
\cite{PHA}, there have been important developments in integrable quantum field theory. Those most
pertinent to this paper are described below in Section \ref{theory}. Amplitude ratios for the universality classes of
the $q$-state Potts model
\cite{dc} and the tricritical Ising model \cite{fms1, fms2} have recently been constructed using field
theoretic approaches. For the Ising model in a magnetic field, transfer matrix results for a lattice
Hamiltonian are compared with $S$-matrix determination of universal amplitudes in \cite{ch}.

Probably better recognized for its connection to the
Ising model in a magnetic field, in
this short paper the dilute A model \cite{WNS} is discussed in relation to the subleading magnetic perturbation of
tricritical Ising model.  That model is introduced in
Section \ref{model} and some quantum field theory results applying to it are outlined in Section \ref{theory}.
In Section \ref{twoone} the dilute A$_3$ model, and some results for it, are described; it is shown to be a
solvable lattice model realization of the $\phi_{2,1}$ perturbation of $\mathcal{M}(4,5)$.  
Motivated by \cite{fms2}, universal amplitude ratios are constructed in
Section \ref{results}. These are found to be in perfect agreement with results from quantum field theory, which
have appeared since the original solution of the dilute A model in \cite{WPSN}. A brief discussion concludes the paper.


\section{The tricritical Ising model}\label{model} 
The tricritical Ising model corresponds to the minimal unitary conformal
field theory $\mathcal{M}(4,5)$ with central charge $c=\case{7}{10}$ and its critical exponents are related to
the conformal weights
\begin{equation}
\Delta^{(5)}_{r,s} =\frac{(5r-4s)^2-1}{80},\qquad 1\leqslant r \leqslant 3,\ 1\leqslant s \leqslant 4.
\label{weights}
\end{equation}

One lattice realization of this model is the Blume-Capel model \cite{BC}, a spin-1 generalization of the Ising
model with vacancies favoured by a chemical potential $\mu$. The Hamiltonian can be written as
\begin{equation}
\mathcal{H}_{\rm{BC}}= -J\sum_{\langle i,j\rangle} s_i s_j -H\sum_{i=1}^N s_i+\mu\sum_{i=1}^N (s_i)^2,
\label{blume}
\end{equation}
where $J$ is the nearest-neighbour coupling and $H$ is an external magnetic field. The lattice variable may be
thought of as  
\begin{equation}
s_i=\sigma_i v_i \qquad \mbox{where} \qquad \sigma_i=\pm 1, \ v_i=0,1. \label{spin}
\end{equation}
The spin is $\sigma_i$ and the occupancy (or vacancy) is expressed
by $v_i$. The model further admits a subleading or staggered magnetic field, in which case it is more properly
called the Blume-Emery-Griffiths model:
\begin{equation}
\mathcal{H}_{\rm{BEG}}= \mathcal{H}_{\rm{BC}}-H_3\sum_{\langle i,j\rangle} v_i v_j(\sigma_i+\sigma_j).
\label{beg}
\end{equation} 
The staggered field favours aligned spins on neighbouring sites.
The phase diagram of this model is like that of He$^3$-He$^4$ mixtures
shown in
\cite{gr}. The features of interest for this paper are its wings of two-phase coexistence, on one of which the phases are
$s_i=0$ and
$s_i=1$, and on the other $s_i=0$ and $s_i=-1$. The subleading magnetic
perturbation of the tricritical Ising model gives access to a one-dimensional section of the full three-dimensional two-phase
coexistence manifold \cite{lmc}.  

\section{ The subleading magnetic perturbation}\label{theory}
Two remarkable advances in quantum field theory should be mentioned in the context of the tricritical Ising
model. Zamolodchikov showed that the minimal conformal field theories $\mathcal{M}(p,p')$ admit
integrable perturbations by one of the operators $\phi_{1,2}$, $\phi_{1,3}$, $\phi_{1,5}$ and $\phi_{2,1}$,
if they are relevant \cite{Z, Za}. It was further established, using the connection of the
quantum field theories to scattering theory, that the minimal unitary theories $\mathcal{M}(p,p+1)$ for $p=3,4,5$ 
perturbed by $\phi_{1,2}$ correspond to the exceptional E$_6$, E$_7$ and E$_8$ Toda field theories \cite{Za,Zb}.

An important study of the mass spectra and phase diagrams of the four relevant perturbations of the tricritical Ising model was
undertaken in \cite{lmc} using the truncated conformal space approach. The two
thermal perturbations are $\phi_{1,2}$ (leading) and $\phi_{1,3}$ (which is related to the chemical potential
$\mu$ in (\ref{blume})). Interestingly, the `usual' magnetic field $H$ in
(\ref{blume}) corresponds to the perturbation $\phi_{2,2}$ of $\mathcal{M}(4,5)$ which is relevant but not
integrable, while the staggered magnetic field in (\ref{beg}) corresponds to the integrable perturbation
$\phi_{2,1}$, though it is not realizable
in a laboratory \cite{gr}.  It is
conceptually and notationally convenient \cite{lmc} to number the fields according to Landau-Ginzburg $\Phi^6$
formulation, from most relevant ($\varphi_1$) to least relevant ($\varphi_4$) on the basis of the associated
conformal weights (\ref{weights}). Table \ref{summ} summarises this and what is to
follow about the manifestations of
these four perturbations.

\Table
{\label{summ} The four perturbations of the tricritical Ising model}
\br
Field& Perturbation&Weight& Lattice model\\
\mr 
$\varphi_1$ Leading magnetic & $\phi_{2,2}$ & $\case{3}{80}$ &  Not integrable\\ \ms
$\varphi_2$ Leading thermal & $\phi_{1,2}$ & $\case{1}{10}$ &  Dilute A$_4$, regime 2\\ \ms
$\varphi_3$ Subleading magnetic & $\phi_{2,1}$ & $\case{7}{16}$ &  Dilute A$_3$, regime 1\\ \ms
$\varphi_4$ Subleading thermal & $\phi_{1,3}$ & $\case{3}{5}$ &  ABF A$_4$, regime III\\
\br
\endTable

The
$S$-matrix for the subleading magnetic
perturbation in particular was studied in \cite{ckm, km}, and using the
thermodynamic Bethe ansatz in \cite{eb}. Mass spectra in general are exactly related to bulk
energies and coupling constants of the appropriate perturbations in
\cite{F}. Degeneracy in the
spectrum for $\phi_{2,1}$ means that only one distinct mass value emerges. 

Other results for the $\phi_{2,1}$ perturbation of $\mathcal{M}(p,p+1)$ can also be found among 
general discussions of integrable perturbed quantum field theories. In particular, expressions for 
one-point correlation functions or vacuum expectation values of the local fields are proposed in
\cite{flzz}. Explicit values are given for the tricritical Ising model, and
compared to numerical results of \cite{GM}. Predictably, two degenerate but asymmetric groundstates are
identified for each sign of the coupling constant.
 
Universal amplitude ratios for the relevant perturbations of the tricritical Ising model have recently been
constructed \cite{fms1, fms2}, by collecting together information from the approaches above, as well as
numerical results. These ratios involve the amplitudes of the correlation lengths $\xi_i$ and of the free energy $f_i$ (where the
subscript $i=1, \ldots, 4$ denotes the field taking the model off-critical)
but are independent of the coupling constants $g_i$ and hence of any nonuniversal metric factors. Further,
they involve the amplitudes of the vacuum expectations of the local fields
\begin{equation}
\langle \varphi_j\rangle _i \equiv B_{ji} \ g_i^{\Delta_j/(1-\Delta_i)}, \label{vev}
\end{equation} 
as well as generalized susceptibilities.
Some of these ratios should be observable both in real-world representations of this
universality class, and the related solvable lattice models, provided the physical quantities required can be
measured or constructed.

Each of the three integrable
perturbations of the tricritical Ising model has been identified with a particular solvable lattice model in its
massive, scaling limit. The elliptic nome of the A$_4$ ABF model \cite{ABF, hu} in regime III corresponds to
the coupling constant of the subleading thermal perturbation, $\phi_{1,3}$.  Smirnov {\cite{smir} conjectured
that lattice models for
$\phi_{1,2}$ and $\phi_{2,1}$  would be based
on RSOS restriction of the Izergin-Korepin \cite{ik} $A_2^{(2)}$ $R$-matrix. The dilute A model \cite{WNS}
is a hierarchy of such models, and provides the two remaining cases.  
Of these, most attention has focused on the dilute A$_4$ model in
regime 2, which is in the universality class of the $\phi_{1,2}$ perturbation, to elucidate
the hidden E$_7$ structure \cite{WP, BS3,js,SB2}. A detailed description of how the dilute A$_3$ model in regime
1 realizes the
$\phi_{2,1}$ perturbation
of the tricritical Ising model will follow in Section \ref{twoone}. Delfino \cite{delf} gives an interesting
phase diagram concerning the points of contact of the regimes of these three solvable models from a discussion
of  double perturbations of the minimal unitary series by $\phi_{1,3}$ and one of $\phi_{1,2}$ or $\phi_{2,1}$.


\section{The dilute A$_3$ model in regime 1 as the tricritical Ising model}{\label{twoone}
The dilute A$_L$ model \cite{WNS} is a hierarchy of solvable restricted solid-on-solid models, defined on the
square lattice and labelled by the allowed number of heights, $L$. The adjacency condition of the model is that
neighbouring sites of the lattice are either occupied by the same height or differ by one.
The face weights of the model are parameterized using elliptic functions. It is solvable in four off-critical
regimes, generated by the elliptic nome $p$ and the value of the crossing parameter, which in regime
1\footnote{This regime labelling differs from \cite{WNS} but is followed in \cite{WPSN} and all subsequent
papers.} is 
\begin{equation*}
\lambda=\frac{\pi}{4}\ \left(\frac{L}{L+1}\right) ,
\end{equation*}
while the central charge, known from equivalence with the O$(n)$ model, for regime 1 is
\begin{equation*}
c=1-\frac{6}{(L+2)(L+1)}.
\end{equation*}

For odd $L$ the elliptic nome breaks the $\mathbb{Z}_2$ symmetry of the underlying adjacency diagram and is
thus magnetic-field like; changing the sign of $p$ causes height relabelling $a \to L+1-a$, where $a=1,
\ldots, L$. It is sometimes necessary to distinguish between regime 1$^+$ where $p>0$ and regime 1$^-$ where
$p<0$.

The singular part of the free energy of the dilute A$_L$ model in regime 1 to leading
order as the nome $p \to 0$ behaves as
\begin{equation*}
f_{\rm s} \sim  \mathcal{A}_L \ p^{\frac{4}{3}\left(\frac{L+1}{L}\right)}, 
\end{equation*}
which follows from the partition function-per-site calculated either using the inversion relation method 
\cite{WPSN} or the largest eigenvalue of the row-to-row transfer matrix \cite{BS1}. For odd $L$ this gave
\cite{WPSN} the critical exponent $\delta=3L/(L+4)$ and the associated scaling dimension
\begin{equation}
\Delta_p=\frac{1}{1+\delta}=\frac{L+4}{4(L+1)}=\Delta^{(L+2)}_{2,1}, \label{delta}
\end{equation}
so that in regime 1 the nome generates perturbation of the unitary minimal model $\mathcal{M}(L+1,L+2)$
by the operator $\phi_{2,1}$. 
This is the basis of the earlier assertion that for $L=3$ the nome relates to the subleading magnetic
perturbation of the tricritical Ising model; or in field theoretic notation, the coupling constant $g_3$
of $\varphi_3$; 
or in the language of the lattice Hamiltonian, the staggered magnetic field. 

Correspondence can also be drawn at the level of details of the groundstates. In regime
1 the A$_3$ model has two ferromagnetic groundstates in the ordered ($|p| \to 1$) limit, for each sign of the elliptic nome
\cite{WPSN}. For one sign of $p$ all sites of the lattice are occupied either by height 1, or by height 2.
Under the field reversal noted before, the groundstates for the other sign of $p$ consist of all lattice sites
being occupied by height 3 or by height 2. The straightforward identification of the dilute
A$_3$ model height variable $a_i$ at site
$i$ with the  spin and vacancy variables (\ref{spin}) via
\begin{equation*}
a_i=2+\sigma_i v_i
\end{equation*}
maps the groundstates to those described in Section \ref{model}. The adjacency condition of the dilute A$_3$ model
then corresponds to each occupied site having as its neighbour either an empty site, or an aligned spin.

Local height probabilities $P^{bc}(a)$, expressing the probability that a site deep in the lattice is occupied
by height
$a$ when the model is in the phase labelled by $(b,c)$ were
calculated for the dilute A$_L$ model ($L$ odd) in
\cite{WPSN}, using the corner transfer matrix technique \cite{Ba}. This
calculation involves considering a series of finite square lattices with boundary sites set to the
configurations ($b,c$), solving recurrence relations and then taking the lattice size to infinity. The
polynomials generated in this process for general $L$ in regime 1 have found application as the required
finitizations of the bosonic side of Rogers-Ramanujan-type identities, termed $\phi_{2,1}$ polynomial
identities \cite{BM}. 

Following Huse
\cite{hu}, generalized order parameters
\begin{equation*}
R_k^{bc}=\sum_{a=1}^{L} \frac{\sin((k+1)a \pi/(L+1))}{\sin(a \pi/(L+1))} \ P^{bc}(a) \qquad k=0, \ldots L-1
\end{equation*}
were defined and the leading order behaviour determined \cite{WPSN}. For $L=3$ in
regime 1$^+$, where the ferromagnetic groundstates have $c=b$,
\begin{equation}
\fl 
R_k^{bb}\sim p^{[(k+1)^2-1]/45} \sin((k+1)\pi s/5)/ \sin(\pi s/5); 
\qquad
s=\cases{1 & b=1\\
3 & b=2. } \label{order}
\end{equation}
The associated scaling dimensions were determined using (\ref{weights}), (\ref{vev}) and (\ref{delta}) to be
\begin{equation*}
\Delta_k=\Delta^{(5)}_{k+1,k+1} \qquad k=1,2.
\end{equation*}
Using the property $\Delta_{r,s}^{(5)}=\Delta^{(5)}_{4-r, 5-s}$, $R_1^{bb}$ is identified with the
expectation value of the operator $\phi_{2,2}$
(or $\varphi_1$) and similarly, $R_2^{bb}$ with $\phi_{1,2}$ (or $\varphi_2$), in both of the two phases
labelled by $b$. (Fairly trivially,
$R_0^{bb}=1$ by definition, and corresponds to $\phi_{1,1}$ or the identity operator.)

The variable $s$ introduced in (\ref{order}) depends only on the height $b$, and provides the identification of the lattice model
groundstates and the phase labelling in \cite{flzz} required if we are to proceed to compare universal quantities. In the field
theory context the phases are 
$|0_s\rangle$ with $s=2, 4$ for positive coupling constant, and $s=1,3$ for negative coupling
constant. (For reasons to do with arbitrary choices made in the Boltzmann weights of the
dilute A model as set up in \cite{WPSN} the elliptic nome is actually $-g_3$, so that regime 1$^+$
corresponds to
$g_3<0$. Applying the appropriate height reversal to the working leading to (\ref{order}) one obtains $s=2,4$ in
regime 1$^-$. Since the perturbation is magnetic, the ratios to be considered in Section \ref{results} are in
fact unchanged by this rather technical detail. A similar observation is made for dilute A$_4$ in
\cite{BS3,SB2}, where
$p<0$ is noted to be the {\em high} temperature regime.)

\section{Universal amplitude ratios} \label{results}
Knowing just $\langle \varphi_j\rangle_3$, $f_{3}$ and $\xi_3$ (see (\ref{corr}) below) does not permit
construction of many of the various amplitude ratios proposed in \cite{fms2}. However, we {\em can} consider the
ratio of $\langle
\varphi_j\rangle_3$ in one phase of the tricritical Ising model (say $s=-1$) with the expectation value of the same operator
in the coexisting phase ($s=0$), for
$j=1,2, 3$:
\begin{equation}
\frac{\langle\varphi_j\rangle_3^{(-)}}{\langle\varphi_j\rangle_3^{(0)}}=\frac{B_{j3}^{(-)}}{B_{j3}^{(0)}}  .\label{ampB}
\end{equation}
From (\ref{vev}), these ratios are pure numbers independent of the nome or $g_3$, and hence of any associated microscopic
scaling factors of the solvable lattice realization of the tricritical Ising model, or of the field theory
considered in
\cite{flzz}.

First, $\langle\varphi_3\rangle_3=-\frac{\partial f_3}{\partial g_3}$,
and since the amplitude of the free energy expression does not depend on the sign of the nome or on the
phase, $B_{33}^{(-)}/B_{33}^{(0)}=1$, in agreement with the identical numerical values given in Table XVII of
\cite{fms2}.

From (\ref{order}) we find
\begin{equation}
\frac{B_{j3}^{(-)}}{B_{j3}^{(0)}}=\frac{R_j^{11}}{R_j^{22}}=\frac{1+\sqrt{5}}{1-\sqrt{5}} \qquad j=1,2. \label{amp}
\end{equation}
In the expressions for $\langle 0_s|\phi_{l,k}|0_s\rangle$ given in \cite{flzz} the only $s$ or phase
dependent factor is
\begin{equation}
{\sin (\pi s |5l-4k|/5)}/{\sin (\pi s/5)}
\end{equation}
which should be compared to the coefficients in (\ref{order}). The amplitude ratios constructed from
the field theory
 are then in perfect accord with (\ref{amp}):
\begin{equation*}
\frac{\langle 0_1|\phi_{1,2}|0_1\rangle}{\langle 0_3|\phi_{1,2}|0_3\rangle}=
\frac{\langle
0_1|\phi_{2,2}|0_1\rangle}{\langle 0_3|\phi_{2,2}|0_3\rangle}=\frac{1+\sqrt{5}}{1-\sqrt{5}}.
\end{equation*}

One further universal amplitude can be calculated. The single correlation length of the model in regime
1 (corresponding to one distinct mass as mentioned in Section \ref{theory})  was calculated from the leading
transfer matrix eigenvalue excitation, constructed by Baxter's exact perturbative method \cite{Ba} using the
Bethe ansatz equations
\cite{BNW}, in
\cite{BS1,BS2}. From the general expression for all $L$ in terms of standard elliptic theta functions (following
the approach taken in
\cite{SB1}),
\begin{equation}
 \xi^{-1}=2 \log \left[\frac{\vartheta_4(\frac{\pi}{12}, p^{\pi/6\lambda})}{\vartheta_4(\frac{5
\pi}{12}, p^{\pi/6\lambda})}\right] =8 \sum_{n=1}^{\infty} \frac{1}{n}\
\frac{(p^{\pi/6\lambda})^n}{1-(p^{\pi/6\lambda})^{2n}}
\sin(\case{n \pi}{2}) \sin(\case{n \pi}{3}),\label{corr}
\end{equation}
we now obtain the coefficient of the leading order term for $L=3$ to be
\begin{equation*}
\xi^{-1}\sim 4 \sqrt{3} p^{{8}/{9}} +\Or\left(\left(p^{{8}/{9}}\right)^3\right).
\end{equation*}
Combining this with the free energy amplitude  $\mathcal{A}_3=2 \sqrt{3}/\cos(4 \pi/9)$ gives the tricritical Ising subleading
magnetic amplitude
\begin{equation}
f_{\rm s} \xi^2 =\frac{1}{8 \sqrt{3} \cos (\frac{4 \pi}{9})}, \label{amp2}
\end{equation}
in agreement with the expression for it in \cite{F}. This universal quantity is related to $R_{\xi}^3$ of \cite{fms2}.


\section{Discussion}
 Only some of the objects used in constructing universal amplitude ratios in \cite{fms2} can be calculated in
the dilute A$_3$ `version' of the tricritical Ising model considered in this paper. To utilize what is available, a different
universal quantity (\ref{ampB}) has been considered. Thus
equations (\ref{amp}) and (\ref{amp2}) go some way towards answering a challenge put out in \cite{fms1}, to independently
determine univeral amplitude ratios for the tricritical Ising model from a solvable lattice formulation.
One barrier to constructing similar ratios to (\ref{ampB}) for the dilute A$_4$ model
in regime 2, which corresponds to the leading thermal perturbation, is that local height probabilities for $L$
even have not yet been calculated (in any regime). 
However, amplitudes akin to
(\ref{amp2}) are discussed for dilute A$_4$ in regime 2 in
\cite{SB2}, where the conjectured
\cite{BS3} E$_7$ mass ratios are confirmed. There are more general observations to be made along the lines of
this paper, and it is intended to report on them in the near future.

\ack
This paper was written while the author enjoyed the hospitality of the
Department of Mathematics and Statistics at the University of Melbourne.
She also acknowledges the importance to it of earlier collaboration on the results re-examined here with Bernard
Nienhuis, Paul Pearce and Ole Warnaar, and recent work with Murray 
Batchelor.

\end{document}